\documentclass[prb,twocolumn,aps,showpacs]{revtex4}
\usepackage{graphicx}% complex graphics
\usepackage{bm}% bold math
\usepackage{amssymb} %math symbols
\begin{document}

\title{The photon absorption edge in superconductors and gapped 1D systems}

\author{V. V. Mkhitaryan$^{1}$, E. G. Mishchenko$^{1}$,
M. E. Raikh$^{1}$, and L. I. Glazman$^{2}$}

\affiliation{$^{1}$Department of Physics, University of Utah,
Salt Lake City, UT 84112\\
$^{2}$Department of Physics, Yale University, New Haven, CT 06520}

\begin{abstract}
Opening of a gap in the low-energy excitations spectrum affects
the power-law singularity in the photon absorption spectrum
$A(\Omega)$. In the normal state, the singularity,
$A(\Omega)\propto [D/(\Omega-\Omega_{\rm th})]^\alpha$, is
characterized by an interaction-dependent exponent $\alpha$. On
the contrary, in the supeconducting state the divergence,
$A(\Omega)\propto (D/\Delta)^\alpha(\Omega-\tilde{\Omega}_{\rm
th})^{-1/2}$, is interaction-independent, while threshold is
shifted, $\tilde{\Omega}_{\rm th}=\Omega_{\rm th}+\Delta$; the
``normal-metal'' form of $A(\Omega)$ resumes at
$(\Omega-\tilde{\Omega}_{\rm th})\gtrsim \Delta\exp(1/\alpha)$. If
the core hole is magnetic, it creates in-gap states; these states
transform drastically the absorption edge. In addition, processes
of scattering off the magnetic core hole involving  spin-flip give
rise to inelastic absorption with one or several {\it real}
excited pairs in the final state, yielding a structure of peaks in
$A(\Omega)$ at multiples of $2\Delta$ above the threshold
frequency. The above conclusions apply to a broad class of
systems, e.g., Mott insulators, where a gap opens at the Fermi
level due to the interactions.
\end{abstract}

 \maketitle

%%%%%%%%%%%%%%%%%%%
\begin{figure}[b]
\centerline{\includegraphics[width=85mm,angle=0,clip]{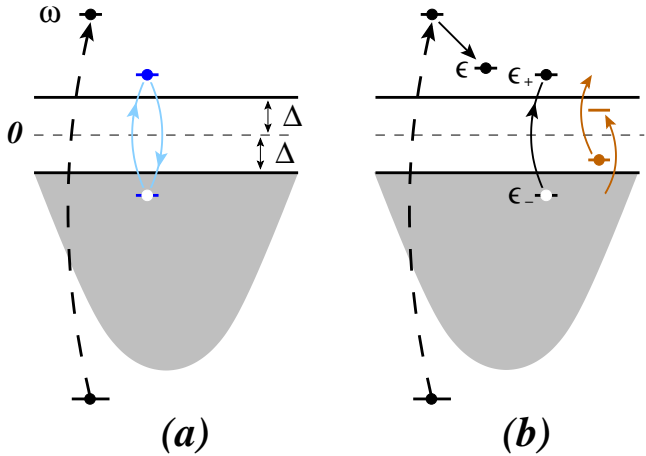}}
\caption{(Color online) Schematic illustration of elastic
absorption, (a), and inelastic absorption, (b). Blue lines
illustrate creation and annihilation of a {\em virtual } pair that
participates in {\it elastic} absorption. Final state of inelastic
absorption is electron with energy $\epsilon$ and a {\it real}
pair, $(\epsilon_+,\epsilon_-)$. Brown lines in (b): since
inelastic absorption is possible {\it only} for a spinful core
hole, in-gap states created by this hole \cite{Rusinov} can also
participate in absorption.} \label{process}
\end{figure}
%%%%%%%%%%%%%%%%%%
\section{Introduction}

It was demonstrated more than $40$ years ago
\cite{Mahan67,Nozieres,review} that electron $x$-ray absorption
coefficient in metal, $A(\omega)$, is strongly modified by
attraction to the localized hole left behind. The threshold
behavior of absorption coefficient was found to be
\begin{equation}
\label{conventional} A(\omega)={\cal
A}_0\left(\frac{D}{\omega}\right)^{\alpha}.
\end{equation}
In Eq. (\ref{conventional}) and thereafter,
$\omega=\Omega-\Omega_{\rm th}$ stands for the difference between
the photon energy and the core-hole energy measured from the Fermi
level, and $D$ is the bandwidth. Prefactor, ${\cal A}_0$, contains
the square of the dipole matrix element between the level and the
conduction band. In the simplest case of a weak short-range
attraction, $V({\bf r})<0$, of electron to the hole
%the shake-up effects can be
%neglected, and
the expression for the exponent $\alpha \ll 1$ has
a form
\begin{equation}
\label{alpha} \alpha=2\nu_0 {\Big |} \int\! d{\bf r}\,V({\bf r})
{\Big |},
\end{equation}
where $\nu_0$ is the density of
states at the Fermi level (we neglect the correction, $-\alpha^2/4$,
originating from the Anderson orthogonality catastrophe, \cite{Nozieres}).
Since the diverging absorption Eq. (\ref{conventional})
comes from all energy scales between $\omega$ and $D$, it is
quite robust. In a finite system, the threshold
behavior depends on additional energy scale, the
level spacing \cite{Baranger}.
%%%%%%%%%%%%%%%%%%%
\begin{figure}[t]
\centerline{\includegraphics[width=90mm,angle=0,clip]{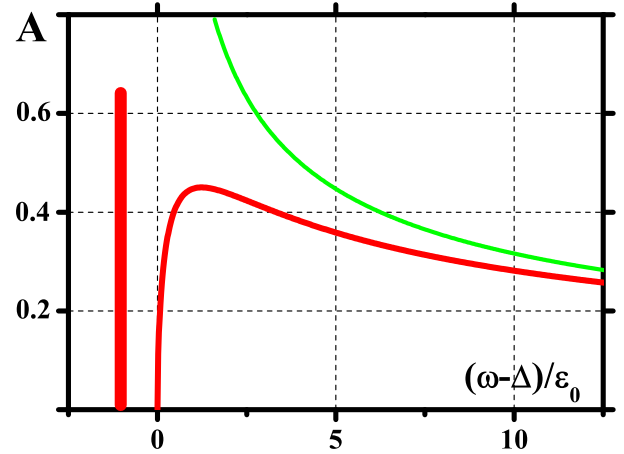}}
\caption{(Color online) Absorption spectrum near the threshold for
spinless (green) and spinful (red) core hole.} \label{graphic}
\end{figure}
%%%%%%%%%%%%%%%%%%

Interest to the singular behavior of $A(\omega)$ near the
threshold got a boost after it was predicted \cite{Matveev92} that
this behavior manifests itself in the resonant-tunneling
current-voltage characteristics. This prediction was later
confirmed in numerous experiments \cite{T1, T2, T3, T4, T5, T6,
T7}. Enhancement of absorption Eq. (\ref{conventional}) was
derived under the assumption that the density of states,
$\nu(\omega)$, is constant $\nu(\omega)=\nu_0$ within the entire
frequency interval, $(-D, D)$. If there is a gap, $2\Delta$, at
the Fermi level the threshold behavior of $A(\omega)$ is singular
even without interaction with a hole:
\begin{equation}
A(\omega)\propto \nu(\omega)=\nu_0
\frac{\omega}{(\omega^2-\Delta^2)^{1/2}}
\end{equation}
and diverges near the edge of the gap. For small $\alpha$ it could
be expected \cite{Ma85} that this strong bare singularity is
weakly affected by the excitonic effects \cite{Mahan67}. Indeed,
the low-energy, $<2\Delta$, many-body processes across the gap,
responsible for Mahan singularity, are suppressed. This reasoning
suggests the form of the absorption in superconductor
\begin{equation}
\label{result1} A(\omega)={\cal
A}_0\left(\frac{D}{\Delta}\right)^{\alpha}\frac{\nu(\omega)}{\nu_0}.
\end{equation}
Eq. (\ref{result1}) crosses over to the conventional behavior
Eq.~(\ref{conventional}) at high frequencies, $\omega$, such that
$\alpha \ln(\omega/\Delta) \sim 1$; in this frequency domain the
effect of superconductivity is negligible, since $\omega \gg
\Delta$.

Even stronger modification of the absorption spectrum takes place,
when the core hole possesses a spin, so that the interaction with
excited electron includes exchange. In this case two new physical
mechanisms come into play. Firstly, a core hole creates in-gap
states \cite{Rusinov} with binding energy $\varepsilon_0 \sim
\alpha^2\Delta$ measured from the edges. These states, in turn,
affect dramatically the elastic scattering of excited electron
transforming the near-gap absorption into
\begin{equation}
\label{magnetic} A(\omega)=\frac{{\cal
A}_0}{\sqrt{2}}\left(\frac{D}{\Delta}\right)^\alpha
\!\frac{\bigl[\Delta(\omega-\Delta)\bigr]^{1/2}}
{(\omega-\Delta)+\varepsilon_0},
\end{equation}
see Fig. \ref{graphic}. The absorption is zero at the threshold
and resumes $(\omega-\Delta)^{-1/2}$ falloff only for
$(\omega-\Delta)\gg \varepsilon_0$. As a "compensation" of the
suppressed absorption, a $\delta$-peak
\begin{equation}
\label{delta} A(\omega)=\frac{{\cal
A}_0}{\sqrt{2}}\left(\frac{D}{\Delta}\right)^\alpha\!\!
\sqrt{\Delta\varepsilon_0}\,\,\delta(\omega-\Delta+\varepsilon_0)
\end{equation}
emerges at the position of the bound state.

There is another {\it many-body} feature in $A(\omega)$, which is
specific for the exchange interaction with core hole. This feature
originates from the fact that exchange interaction of electron
with localized magnetic impurity in metal can be accompanied by
creation of an electron-hole pair  \cite{KamGlaz}. The underlying
reason is that localized spin emerges as a result of the on-site
Hubbard repulsion of two electrons. On the other hand, with
electron-electron interaction, {\em two electrons} can be excited
by a {\it single photon} \cite{MishchGlazReiz,DaiRaikhShah}. In
the presence of a rigid superconducting gap, this process starts
from the threshold \cite{MkhitRaikh} $\omega=\omega_1=3\Delta$,
which corresponds to {\em inelastic} absorption with electron and
additional pair in the final state. This process is schematically
illustrated in Fig. \ref{process}b. More additional pairs in the
final state give rise to anomalies at
$\omega=\omega_n=(2n+1)\Delta$, which have the form
\begin{equation}
\label{result2} \frac{\delta A(\omega)}{A(n\Delta)}\sim
\alpha^{2n}\left(\omega-\omega_n\right)^{n-1/2}
\theta(\omega-\omega_n).
\end{equation}

\section{Derivation of Eq. (\ref{result1})}
\subsection{Time dependent superconducting Green functions}

An efficient way \cite{Nozieres} to derive Eq.
(\ref{conventional}) is to consider scattering of excited electron
by a transient potential, $V({\bf r}) \theta(t)$, and perform
calculation in the time representation. In this representation the
Green function of the normal metal $G_0(t)=\int d\omega e^{i\omega
t}\sum_q1/(\omega-\xi_q\pm i0)$ ($+$ or $-$ depending on
$\text{sgn}(\xi_q)$) has the form
$G_0(t)=-\nu_0\left(t-iD^{-1}\text{sgn}(t)\right)^{-1}$, where $D$
is the bandwidth. Generalization of the scattering approach to
superconductor requires the time representation of the
superconducting single-particle Green function
\begin{equation}
\label{Greenfunction} \hat{G}(\omega,q)=
\frac{\hat{\Lambda}_{+}(q)}
{\omega-\epsilon_q+i0}+\frac{\hat{\Lambda}_{-}(q)}
{\omega+\epsilon_q-i0},
\end{equation}
where $\epsilon_q=\sqrt{\xi_q^2+\Delta^2}$ is the spectrum of
superconductor; $\xi_q=v_{\scriptscriptstyle F}q$ with
$q=(k-k_{\scriptscriptstyle F})$ being the momentum measured from
the Fermi momentum, $k_{\scriptscriptstyle F}$, and
$v_{\scriptscriptstyle F}$ is the Fermi velocity. The projection
operators $\hat{\Lambda}_{\pm}(q)$ are $2\times 2$ matrices
\begin{equation}
\label{Lambda}
\hat{\Lambda}_{\pm}(q)=\frac12\left(\begin{array}{cc}
1\pm\frac{\xi_q}{\sqrt{\xi_q^2+\Delta^2}}&
\mp\frac{\Delta}{\sqrt{\xi_q^2+\Delta^2}}\\
\mp\frac{\Delta}{\sqrt{\xi_q^2+\Delta^2}}&
1\pm\frac{\xi_q}{\sqrt{\xi_q^2+\Delta^2}}
\end{array}\right),
\end{equation}
with  following properties:
$\hat{\Lambda}^2_{\pm}(q)=\hat{\Lambda}_{\pm}(q)$ and
$\hat{\Lambda}_{+}(q)+\hat{\Lambda}_{-}(q)=1$. In the basis of
eigenfunctions of the Bogoliubov-de Gennes Hamiltonian,
interaction with the short-range potential is described by the
diagonal matrix
\begin{equation}
\label{matrixV} V_q= -\frac\alpha{2\nu_0}\hat{V};\quad \hat{V}=
\left(\begin{array}{cc}
1& 0\\
0& -1
\end{array}\right).
\end{equation}
Note that time-dependent $2\times 2$ Green function of a
superconductor, obtained as a result of integration $d\omega
e^{i\omega t}$ of Eq. (\ref{Greenfunction}), and subsequent
summation over momentum, $q$, can be conveniently expressed in
terms of zeroth and first-order Bessel functions, namely
\begin{equation}
\label{hatG} \hat{G}(t)=\left(\begin{array}{cc}
G(t)& F(t)\\
F(t)& G(t)
\end{array}\right),
\end{equation}
where the normal and anomalous Green functions, $G(t)$ and $F(t)$,
are given by
\begin{eqnarray}
\label{timeGf}
&&G(t){\Big |}_{Dt>1}=
\frac{\pi\Delta\nu_0}{2}\text{sgn}(t)\Bigl[iJ_1(\Delta \vert t\vert)+
Y_1(\Delta \vert t \vert)\Bigr],\qquad\\
&&F(t){\Big |}_{Dt>1}= -i\frac{\pi\Delta\nu_0}{2}\Bigl[iJ_0(\Delta
\vert t\vert)+ Y_0(\Delta \vert t \vert)\Bigr].\label{timeFf}
\end{eqnarray}
In the limit $\Delta \rightarrow 0$ the normal-metal Green
function, $G_0(t)$, is recovered from   Eq. (\ref{timeGf}) by
using the small-$t$ asymptote  $Y_1(\Delta t)\approx -2/(\pi
\Delta t)$, while $F(t)\rightarrow 0$.
%%%%%%%%%%%%%%%%%%%
\begin{figure}[t]
\centerline{\includegraphics[width=70mm,angle=0,clip]{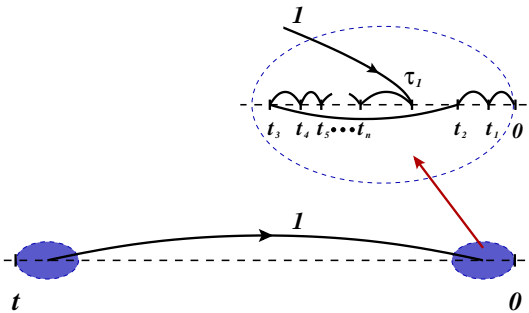}}
\caption{(Color online) Conventional arrangement of times
\cite{Nozieres} in $n$- fold integral Eq. (\ref{Ln}) describing
contribution to the response function due to $n$ successive
scatterings by the core hole. Time intervals, $|t_i-t_{i+1}|$, are
distributed unevenly; central interval corresponding to line $1$
is only slightly smaller than $|t|$. Remaining intervals contained
in the boundary ellipses are $\ll |t|$.  Inset: blowup of the
right end of the line~$1$.} \label{insertion}
\end{figure}
%%%%%%%%%%%%%%%%%%

\subsection{Shape of the absorption edge}

In superconductor, we generalize the response function
to a $2\times 2$ matrix, $\hat{L}(t)$, so that the absorption
coefficient is given by the diagonal matrix  element
\begin{equation}
\label{expression}
A(\omega)= \frac{{\cal A}_0}{\pi \nu_0}\,\text{Re}
\int^0_{-\infty}
\!\!\!dt\,
\exp(-i\omega t)\left[\hat{L}(t)\right]_{11}.
\end{equation}
As a result of matrix generalization, the expansion of the
response function in powers of $\alpha$,
\begin{equation}
\hat{L}(t)=\sum_n\left(-\frac\alpha{2\nu_0}\right)^n\hat{L}_n(t),
\end{equation}
has the $2\times 2$ coefficients, $\hat{L}_n(t)$, which are given
by the following $n$-fold integrals \cite{Nozieres,Matveev92} of
the single-particle Green function, $\hat{G}(t)$,
\begin{equation}
\label{Ln} \hat{L}_n(t)=i\!\int_t^0\!\!\!dt_1\cdot\cdot\!
\int_t^0\!\!\!dt_n \hat{G}(-t_1)\hat{V} \hat{G}(t_1-t_2)\hat{V}
\cdot\cdot\hat{V}\hat{G}(t_n-t).
\end{equation}
In the normal metal, evaluation of $A(\omega)$ is based on exact
analytical result \cite{Nozieres} for the infinite sum
\begin{eqnarray}\label{exact}
&&\hspace{-0.5cm}\sum_{n=0}^{\infty}\!\left(\!-\frac{\alpha}{2\nu_0}\!\right)^{\!\!n}\!\!\!
\int_t^0\!\!\!dt_1\cdots \!\int_t^0\!\!\!dt_n
G_0(\tau\!-t_1)\cdots
%G_0(t_1\!-t_2)\cdot\cdot
G_0(t_n\!-\tau^\prime)
\nonumber\\
&&\hspace{-0.5cm}=G_0(\tau-\tau^\prime)\left[\frac{(t-\tau)\tau^\prime}
{(t-\tau^\prime+iD^{-1})(\tau+iD^{-1})}\right]^{\alpha/2}.
\end{eqnarray}
To arrive to Eq. (\ref{conventional}) one has to set $\tau =0$ and
$\tau^{\prime}=t$, after which the square bracket in  Eq.
(\ref{exact}) reduces to $(-iDt)^\alpha$, and integrate
$dt\exp(-i\omega t)$. Characteristic times $t_i$ in the relation
Eq. (\ref{exact}) are arranged unevenly as illustrated in Fig.
\ref{insertion}. The central interval is $\approx t$, so that
$t_i$ are located in the close proximity, $\tau_1$ or $\tau_2$
(see  Fig. \ref{insertion}) either to $0$ or to $t$.
%{\bf within $t\exp(-1/\alpha)$}.
It is important that in superconducting case the arrangement
remains the same, and moreover, as we will see, $\Delta\tau_1$ and
$\Delta\tau_2$ are always $\ll 1$. This means that
$\hat{G}(t_i-t_{i+1})$ can be replaced by $G_0(t_i-t_{i+1})$ times
the unit matrix. As a result, the matrix structure of $\hat{V}$
drops out. The only Green function that retains the matrix
structure is $\hat{G}(\tau_1+\tau_2-t)$, Fig. \ref{insertion}.
However, in the component $\hat{L}_{11}$, the anomalous Green
function drops out, so that
\begin{equation}
\label{leL} \hat{L}_{11}(t)\simeq
i(iD)^\alpha\,\alpha^2\!\!\!\!\!\!\!
%\int\limits_t^0\!\!\frac{d\tau_1}{|\tau_1|^{1-\alpha/2}}\!
%\int\limits_t^0\!\! \frac{d\tau_2}{|\tau_2|^{1-\alpha/2}}\,
\int\limits_{0\atop{\tau_1+\tau_2\leq |t|}}^{|t|}\!\!\!\!
\frac{d\tau_1d\tau_2}{(\tau_1\tau_2)^{1-\alpha/2}} \,
G(-\tau_1-\tau_2-t),
\end{equation}
where $G(t)$ is defined by Eq. (\ref{timeGf}). Eq. (\ref{result1})
immediately follows from Eqs. (\ref{leL}) and  Eq.
(\ref{expression}).  The Green function $G$ in Eq. (\ref{leL})
generates the density of states, $\nu(\omega)$, in Eq.
(\ref{result1}).
%Substituting  Eq. (\ref{largeasympt} ) into Eq. (\ref{leL}) and then Eq. (\ref{leL})
%into Eq. (\ref{expression}), we recover the result Eq. (\ref{result1}).
One point should be clarified with regard to the validity of
the above result Eq. (\ref{leL}). We used the
normal-metal solution Eq. (\ref{exact}).
This is justified since integrals over $\tau_1$, $\tau_2$
in Eq. (\ref{leL}) come from $\tau_1,\tau_2 \sim \Delta^{-1}\exp(-1/\alpha)$.
This also validates the assumption $\Delta\tau_1, \Delta\tau_2 \ll 1$, which
we used to disregard the matrix structure of $\hat{G}(t_i-t_{i+1})$.
%%%%%%%%%%%%%%%%%%%%%%%%%%%%%%%%
\begin{figure}[b]
\centerline{\includegraphics[width=70mm,angle=0,clip]{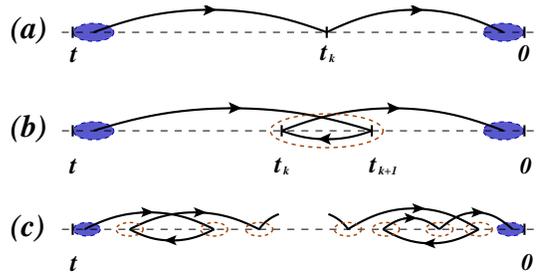}}
\caption{(Color online) Examples of "unconventional" time domains
in the integrand of Eq. (\ref{Ln}); (a): Position of the point,
$t_1$, such that $|t_1|\gg\Delta^{-1}$, $|t-t_1|\gg\Delta^{-1}$,
does not contribute to $\hat{L}_n$ by virtue of Eq. (\ref{3mp});
(b): As long as $|t_1|\gg\Delta^{-1}$, $|t-t_1|\gg\Delta^{-1}$,
and $|t_2|\gg\Delta^{-1}$, $|t-t_2|\gg\Delta^{-1}$, contribution
of the arrangement of times vanishes upon integration over $t_1$
or $t_2$, see Eq. (\ref{zero2}); (c): For the same reason, "long"
($\gg\Delta^{-1}$) intervals in the general "unconventional"
arrangement yield vanishing contribution to $\hat{L}_n$, and thus
to the absorption at the threshold, $(\omega-\Delta)\ll\Delta$.
%Arrangement (b) with $|t-t_1|\lesssim\Delta^{-1}$ and
%$|t_2|\lesssim\Delta^{-1}$, i.e., the path $0\rightarrow
%t\rightarrow0\rightarrow t$, is responsible for $3\Delta$ anomaly.
} \label{diag1}
\end{figure}
%%%%%%%%%%%%%%%%%%

\subsection{Unconventional arrangements of times}

There still remains a question whether or not the matrix structure
of the superconducting Green functions, which becomes important
near the threshold $(\omega-\Delta)\ll\Delta$, gives rise to the
contributions to $A(\omega)$, caused by "unconventional"
arrangements of times, $t_i$, ($|t_i|\gg\Delta^{-1}$), as shown in
Fig. \ref{diag1}a and Fig. \ref{diag1}b; these arrangements are
not relevant in the normal-metal case. For example, the simplest
such "unconventional" arrangement, Fig. \ref{diag1}a, manifests
itself as an extra combination
\begin{equation}
\label{fig1a}\int_t^0\!\!\!dt_k \,\hat{G}(t_{k-1}-t_k)\hat{V}
\hat{G}(t_k-t_{k+1})
\end{equation}
in the integrand Eq. (\ref{Ln}). Since the arguments of $\hat{G}$
in Eq. (\ref {fig1a}) are large, one can use the long-time
asymptote
\begin{equation}
\label{largetG} \hat{G}(t){\Big |}_{\Delta |t|\gg 1}\approx
G_S(t)\left(\begin{array}{cc}
1&-\text{sgn}(t)\\
-\text{sgn}(t)&1
\end{array}\right),
\end{equation}
where the $G_S(t)$ is the $\Delta t \gg 1$ asymptote of Eq.
(\ref{timeGf})
\begin{equation}
\label{largeasympt} G_S(t)= \nu_0\,\text{sgn}(t)
\left(\frac{\pi\Delta}{2\vert t
\vert}\right)^{1/2}\!\!i\,e^{-i\Delta|t| +3\pi i/4 }.
\end{equation}
Note however, that the matrix structure in the integrand of Eq.
(\ref{fig1a}) is
\begin{equation}
\label{3mp} \left(\begin{array}{cc}
1&-1\\
-1&1 \end{array}\right) \left(\begin{array}{cc}
1&0\\
0&-1
\end{array}\right)
\left(\begin{array}{cc}
1&-1\\
-1&1
\end{array}\right)=0,
\end{equation}
Thus, we turn to the next possible arrangement of times Fig.
\ref{diag1}b; the corresponding combination in Eq. (\ref{Ln})
coming from this arrangement reads
\begin{equation}
\label{fig1b} \int_t^0\!\!\!dt_k\! \int_t^0\!\!\!dt_{k+1}
\hat{G}(t_{k-1}-t_k)\hat{V}\hat{G}(t_k-t_{k+1})\hat{V}
\hat{G}(t_{k+1}-t_{k+2}).
\end{equation}
%%%%%%%%%%%%%%%%%%%
\begin{figure}[t]
\centerline{\includegraphics[width=80mm,angle=0,clip]{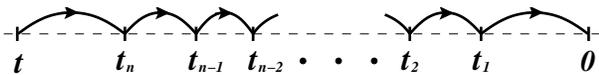}}
\caption{
%(Color online)
For exchange interaction with the core hole, "unconventional"
arrangement of times, $(t_i-t_{i+1})\gg\Delta^{-1}$, dominates the
near-threshold, $(\omega-\Delta)\ll\Delta$ absorption. Odd $n$
describes the absorption peak at $\omega=\Delta-\varepsilon_0$.}
\label{magneticpic}
\end{figure}
%%%%%%%%%%%%%%%%%%
To integrate over $t_k$, we perform multiplication of the first
three matrices and obtain
\begin{widetext}
\begin{equation}
\label{factor} G_S(t_{k-1}-t_k)\!\left(\!\begin{array}{cc}
1& -1\\
 -1& 1
\end{array}\!\right)\!\!
\!\left(\!\begin{array}{cc}
 1& 0\\
 0& -1 \end{array}\!\right)\!\!
\!\left(\!\begin{array}{cc}
G(t_{k}-t_{k+1})&  F(t_{k}-t_{k+1})\\
F(t_{k}-t_{k+1})&  G(t_{k}-t_{k+1})
\end{array}\!\right)
=G_S(t_{k-1}-t_k)\bigl[G(t_{k}-t_{k+1})+ F(t_{k}-t_{k+1})\bigr]
\left(\!\begin{array}{cc}
 1& 1\\
 -1& -1 \end{array}\!\right).
\end{equation}
\end{widetext}
Then the integration over $t_k$ in Eq. (\ref{fig1b}) reduces to
\begin{equation}\label{zero1}
\int\limits^{t_{k-1}-t_{k+1}}_{t_{k+2}-t_{k+1}}\!\! d\tau
\bigl[G(\tau)+ F(\tau)\bigr]G_S(t_{k-1}-t_{k+1}-\tau),
\end{equation}
where we introduced a variable $\tau =t_k-t_{k+1}$. Typical
distance between the points, $|t_{k-1}-t_{k+1}|$ and
$|t_{k+1}-t_{k+2}|$, is $\gg\Delta^{-1}$, which suggests that the
limits of integration can be extended to $\pm\infty$. Upon this
extension we get
\begin{equation}\label{zero2}
\frac{e^{i\Delta (t_{k+1}-t_{k-1})}} {\sqrt{|t_{k+1}-t_{k-1}|}}
\int_{-\infty}^{\infty}\!\!\! d\tau\bigl[ G(\tau)+ F(\tau)\bigr]
e^{i\Delta\tau},
\end{equation}
which is identical zero. The same reasoning rules out
\cite{footnote1} the more complex "unconventional" arrangements of
times {\it at the threshold}, like the ones shown in Fig.
\ref{diag1}c. These arrangements, however, become essential in the
case of exchange interaction with core hole, to which we now turn.

\section{Exchange interaction with core hole}

Exchange interaction with core hole corresponds to replacement
\begin{equation}
V({\bf r})\rightarrow J\delta({\bf r})\left({\bf S}\cdot{\bm
\sigma}\right),
\end{equation}
where ${\bf S}$ is a localized spin, and ${\bm \sigma}$ is
electron spin operator. To illustrate the dramatic impact which
the exchange interaction has on the near-threshold absorption, we
return to Fig. \ref{diag1}a and corresponding expression Eq.
(\ref{fig1a}). For potential interaction with core hole, this
expression was identical zero by virtue of relation Eq.
(\ref{3mp}). Recall now that in the stationary problem the
diagonal part of the exchange interaction, $V({\bf
r})S^z\sigma^z$, creates two in-gap bound states \cite{Rusinov}:
one below the upper edge by
\begin{equation}
\varepsilon_0=\frac{\pi^2\alpha^2\Delta}8,
\end{equation}
and one above the lower edge by $\varepsilon_0$. The reason behind
this effect is that $S^z\sigma^z$ effectively transforms the
operator $\hat{V}$ in Eq. (\ref{matrixV}) into the unity matrix.
An immediate consequence of this transformation for our
calculation is that the contribution Eq. (\ref{fig1a}) becomes
{\it finite}. Subsequently, the contribution Fig. \ref{diag1}b and
all higher-order "unconventional" contributions illustrated in
Fig. \ref{magneticpic}a are also finite. Within our formalism, the
in-gap bound states emerge as poles,
$1/[\omega\pm(\Delta-\varepsilon_0)]$, of the Green function upon
summation \cite{footnote} of infinite series of diagrams.
%with general term depicted in Fig. \ref{magneticpic}a.

In deriving Eq. (\ref{magnetic}) for $A(\omega)$ near the
threshold, we in fact repeat all the steps which would render the
stationary in-gap states. Namely, we notice that the phase
$\Delta\sum_k\vert t_{k+1}-t_k \vert$ of the integrand in Eq.
(\ref{Ln}) is {\em large}, which insures that the dominant
contribution to $L_n(t)$ comes from the domain $0<t_1 < t_2 \cdots
<t$, see Fig. \ref{magneticpic}a, when the net phase is $\Delta
t$; contributions from the domains where $t_m$ are not ordered are
suppressed by oscillations of the integrand. Thus we conclude that
the integral Eq.~(\ref{Ln}) is dominated by $t_m\sim t(m/n)$. For
the asymptote Eq.~(\ref{largeasympt}) to be applicable in this
domain, the condition $(t_{m+1}-t_m)\sim t/n \gg \Delta^{-1}$ must
be met. With $t_m$ ordered, the $n$-fold integration in
Eq.~(\ref{Ln}) can be carried out with the help of the identity
\begin{equation}
\label{used} \int_a^b\!\frac{dx}{\sqrt{(x-a)(b-x)}}=\pi.
\end{equation}
Depending on the parity of $n$, the remaining integration, upon
introducing the variables $z_i=t_i/t$, reduces
%either
to
\begin{equation}
\int\limits_0^1dz_1\int\limits_{z_1}^1dz_2 \cdots\!\!\!
\int\limits_{z_{(n-3)/2}}^1
\!\!\!dz_{(n-1)/2}=\frac{1}{\Gamma\left(\frac{n+1}2\right)}
\end{equation}
for odd $n$, or to
%, or to
\begin{equation}
\int\limits_0^1dz_1\int\limits_{z_1}^1dz_2
\cdots\!\!\!\int\limits_{z_{n/2-1}}^1
\!\!\!dz_{n/2}(1-z_{n/2})^{-1/2}
=\frac{\sqrt{\pi}}{\Gamma\left(\frac{n+1}2\right)}
\end{equation}
for even $n$. Finally we get
\begin{equation}
\label{Lanswer}
\left[\hat{L}_n(t)\right]_{11}=(-1)^n\left(\frac{\pi^2\nu_0^2\Delta}2
\right)^{\frac{n+1}{2}} \frac{(-it)^{\frac{n-1}{2}}}
{\Gamma\left(\frac{n+1}{2}\right)}\,e^{i\Delta t}.
\end{equation}
The product, $\alpha^n\hat{L}_n(t)$, has a sharp maximum at
$n\sim\alpha^2\Delta t$, so that $\Delta t/n\sim1/\alpha^2$ is
large, which justifies the above assumption $(t_{m+1}-t_m)\gg
\Delta^{-1}$.

The sum over even $n$, $\hat{L}_{even}(t) = \sum_{even}
(\alpha/2\nu_0)^n\hat{L}_n(t)$, leads to the result
Eq.~(\ref{magnetic}). Most conveniently it can be seen by
transforming to the frequency domain, since the expansion of
Eq.~(\ref{magnetic}) in powers of $\alpha^2$ has a form
\begin{equation}\label{expand}
A(\omega)={\cal
A}_0\left(\frac{\Delta}{2}\right)^{\frac{1}{2}}\sum_{p=0}^\infty
\left(\frac{\pi^2\alpha^2\Delta}8\right)^p
\frac{(-1)^p}{(\omega-\Delta)^{p+\frac{1}{2}}}.
\end{equation}
This expansion coincides term by term with  the sum,
\begin{equation}
{\cal A}_0\sum_{p}\left(-\frac\alpha{2\nu_0}\right)^{2p}
\int^0_{-\infty}\!dt\,
\left[\hat{L}_{2p}(t)\right]_{11}\exp(-i\omega t),
\end{equation}
with $\hat{L}_{2p}(t)$ given by Eq. (\ref{Lanswer}). The sum over
odd terms results in a simple exponent,
\begin{equation}\label{odd}\hat{L}_{odd}(t)=
\sum_{odd} \left(-\frac\alpha{2\nu_0}\right)^n\hat{L}_n(t) \propto
\exp[i(\Delta-\varepsilon_0) t].
\end{equation}
This exponent gives rise to the $\delta$-peak, Eq.~(\ref{delta}),
in the absorption spectrum.

\section{Inelastic absorption}

Up to now we neglected the spin-flip part,
\begin{equation}
J\delta({\bf r})[S^+\sigma^- +S^-\sigma^+],
\end{equation}
of the exchange interaction. As it was mentioned in the
Introduction, this spin-flip part of interaction between electron
and core hole creates an effective electron-electron scattering
\cite{KamGlaz}. This explains the possibility of inelastic
processes with three quasiparticles in the final state, as
illustrated in Fig. \ref{process}b. The threshold of inelastic
process is $\omega=3\Delta$. Here we will restrict ourself only to
the behavior of inelastic absorption away from the threshold,
$(\omega-3\Delta)\gg\varepsilon_0$, and follow the calculation in
Ref. \cite{MkhitRaikh}. A great simplification away from threshold
is that a "golden-rule"- based calculation is sufficient. The rate
of the process depicted in Fig. \ref{process}b is given by the
following sum over the quasiparticle states with energies,
$\epsilon$, $\epsilon_+$, and $\epsilon_-$,
\begin{equation}
\label{rate} W(\omega)=2\pi\sum_{\epsilon, \epsilon_+, \epsilon_-}
{\Big |}\frac{\alpha_{sf}}{\omega-\epsilon}{\Big |}^2
\delta(\omega-\epsilon- \epsilon_++ \epsilon_-),
\end{equation}
where the first factor is the square of the amplitude, which is
non-zero since the process involves  a spin-flip \cite{KamGlaz},
and the dimensionless spin-flip coupling constant is
\begin{equation}
\alpha_{sf}=J\nu_0\sqrt{S(S+1)}.
\end{equation}
Near the threshold, $\omega=3\Delta$, we have
$\epsilon\approx\Delta$, $\epsilon_+\approx\Delta$, and
$\epsilon_-\approx-\Delta$. The matrix element near the threshold
is approximately constant. This simplifies the summation in Eq.
(\ref{rate}) to
\begin{eqnarray}
\label{inabs} &&W(\omega)= \frac{\pi\,\alpha_{sf}^2}{2\Delta^2}\\
&&\times\!\int\limits_\Delta^\infty\!\! d\epsilon\,\nu(\epsilon)
\int\limits_\Delta^\infty\!\! d\epsilon_+\nu(\epsilon_+)\!\!
\int\limits^{-\Delta}_{-\infty}\!\! d\epsilon_-\nu(\epsilon_-)\,
\delta(\omega-\epsilon- \epsilon_++ \epsilon_-),\nonumber\\
&&=\frac{\pi^2\alpha_{sf}^2}2
\left(\frac{\omega-3\Delta}{2\Delta}\right)^{1/2}.\nonumber
\end{eqnarray}
Note that in the close vicinity of the threshold,
$|\omega-3\Delta|\lesssim \varepsilon_0$, in-gap states created by
the spinful core hole participate in the absorption, as
illustrated in Fig. \ref{process}b. Namely, a pair of
quasiparticles in the final state can consist, e.g., of one
quasiparticle excited above the gap and empty lower in-gap state.
\section{Discussion}
%\subsection{Gapped 1D systems}
Our results Eqs. (\ref{result1}), (\ref{magnetic}) establish the
threshold behavior of $A(\omega)$ for a general situation when the
density of states is strongly modified near the Fermi level but
assumes a constant value away from the Fermi level. A notable
example is a 1D {\em interacting} system. The shape of the
Fermi-edge singularity in 1D interacting electron gas in the
Luttinger-liquid regime has been studied in
\cite{GogolinKaneGlazman} using the bosonization technique.
Backscattering plays an important role in the exponent of the
absorption. When backscattering opens a gap, the physics described
in the present paper comes into play. The case of 1D Mott
insulator near half filling makes the behavior of $A(\omega)$ even
richer, since the doping shifts the threshold. A related example
is the Peierls insulator, when the charge density wave and ensuing
gap at the Fermi level are due to electron-phonon interactions.
Note, that in the latter case the gap is orders of magnitude
larger than in superconductor.

Speaking about conventional setting for Fermi-edge absorption in
metals, singularity in $A(\omega)$ is smeared due to the finite
lifetime, $\gamma$, of the core hole. In our consideration we
assumed that the gap, $2\Delta$, exceeds $\gamma$. In most
experiments in metals the smearing of the edge is a fraction of
eV, i.e., much bigger than a typical $2\Delta$-value. However, the
origin of this smearing is not a natural core hole lifetime
broadening but rather a finite instrumental resolution
\cite{LT79}. The fact that observed absorption shape is a
convolution of the singular $A(\omega)$, a Gaussian, which is
measurement-related, and a Lorentzian, describing natural core
hole lifetime, allows to separate the two contributions to the
edge smearing. Early attempts \cite{LT77} of such separation
yielded $\gamma=40$ meV for $2p$ core hole. In the other
experiment \cite{LT78} involving core hole four times shallower
than in Ref. \onlinecite{LT77}, the natural width was found to be
four times smaller, $\gamma=10$~meV. In later experiment
\cite{LT90}, where the full broadening, $29$~meV, was very small,
analysis of the data for the same absorption line as in
Ref.~\onlinecite{LT78} revealed even smaller value of the core
hole width in simple metals, $\gamma= 4$~meV.

As a final remark, the relevance of the exchange interaction of
electron with the core hole was first pointed out in Ref.
\onlinecite{Girvin}.

%\subsection{Tunneling via quantum dot}
%Resonant-tunneling experiments \cite{T1, T2, T3, T4, T5, T6, T7},
%carried out in the structures with normal-metal leads, confirm the
%prediction \cite{Matveev92} that Fermi-edge singularity manifests
%itself in the $I$-$V$ curves. In this regard, we predict that the
% $I$-$V$ curve in the structure with superconducting leads has  either
%the form Eq. (\ref{result1}) or Eq. (\ref{magnetic}) with $\omega$
%replaced by the bias, $eV$, depending on whether the resonance
%localized state left behind is spinless or spinful.
%with which an empty localized state scatters electron in the lead.
%Note that, since superconducting  gap shrinks with temperature,
%the crossover between the regimes Eq. (\ref{magnetic}) and Eq.
%(\ref{result1}) can be achieved simply by increasing temperature
%or applying magnetic field.
%Note that the first experimental studies of electron transport
%through quantum dot structures based either on InAs \cite{Exp1,
%Exp2} or carbon nanotubes \cite{Tubes1, Tubes2, Tubes3} with
%superconducting leads were carried out only recently.

\section{Acknowledgments}
We acknowledge the hospitality of the KITP at UCSB, where research
was supported in part by the National Science Foundation under
Grant No. PHY05-51164. We also acknowledge the support of DOE
grant No. DE-FG02-06ER46313 (EM), Petroleum Research Fund grant
No. 43966-AC10 (VM and MR), and NSF grant DMR-0906498 (LG).


\begin{references}
\bibitem{footnote} More precisely, to recover stationary in-gap states, one has to
perform $(-\infty, \infty)$ integrations instead of limited
integration intervals in Eqs. (\ref{expression}) and (\ref{Ln}).
\end{references}
\end{document}